\documentclass[aps,pre,twocolumn,floatfix,showpacs]{revtex4}
\usepackage[dvips]{graphicx}

\begin{document}

\title{The Behavior of Granular Materials under Cyclic Shear}
\author{Nathan W. Mueggenburg} \affiliation{University of Chicago
Department of Physics and James Franck Institute} \date{\today}


\begin{abstract}
The design and development of a parallel plate shear cell for the
study of large scale shear flows in granular materials is presented.
The parallel plate geometry allows for shear studies without the
effects of curvature found in the more common Couette experiments.  A
system of independently movable slats creates a well with side walls
that deform in response to the motions of grains within the pack.
This allows for true parallel plate shear with minimal interference
from the containing geometry.  The motions of the side walls also
allow for a direct measurement of the velocity profile across the
granular pack.  Results are presented for applying this system to the
study of transients in granular shear and for shear-induced
crystallization.  Initial shear profiles are found to vary from
packing to packing, ranging from a linear profile across the entire
system to an exponential decay with a width of approximately $6$ bead
diameters.  As the system is sheared, the velocity profile becomes
much sharper, resembling an exponential decay with a width of roughly
$3$ bead diameters.  Further shearing produces velocity profiles which
can no longer be fit to an exponential decay, but are better
represented as a Gaussian decay or error function profile.  Cyclic
shear is found to produce large scale ordering of the granular pack,
which has a profound impact on the shear profile. There exist periods
of time in which there is slipping between layers as well as periods
of time in which the layered particles lock together resulting in very
little relative motion.
\end{abstract}

\maketitle

\section{Introduction}

A static packing of granular materials can support a nonzero shear
stress through a highly inhomogeneous network of forces
\cite{degennes1999}.  Upon increasing the shear stress, the granular
material yields and starts to flow.  Yet the inhomogeneous force
network remains, leading to unusual flow properties
\cite{jaeger1996,jaeger1996rmp}.  The characteristics of the
initiation of this flow, and the properties of the flow, have
implications for industry \cite{ennis1994,knowlton1994}.  An
understanding of these flows and the nature of the transition between
static and flowing behavior holds promise for making connections to
other non-equilibrium systems \cite{liu1998}.

In the regime of rapid dilute flows, in which there is a continuous
input of energy into the system, granular dynamics are dominated by
short-lived collisions between particles.  In many ways, these flows
are similar to those in a molecular gas and may be described by a
granular kinetic theory \cite{jenkins1983}.  In contrast, when a
granular material is dense and slow moving, then individual grains
are in contact with several other grains at once, and these contacts
persist for extended periods of time.  In this regime, the initiation
of shear necessitates dilation of the pack.  In order for individual
grains to move past each other, they must lower the local packing
density.  In confined systems, as is often the case in two-dimensional
studies, the global packing fraction of particles is fixed, but
locally the packing fraction may vary extensively in response to shear
\cite{veje1999}.  In some geometries, the presence of a free surface
allows the packing to adjust its density dynamically to the applied
shear \cite{torok2000,bocquet2002}.  This dilation of the granular
pack has a profound impact on the shear flows.

Numerous systems have been created for studying the flow of various
types of granular materials (see reference \cite{schwedes2003} for a
review of granular flow testers).  The details of the flowing state
have been studied in a variety of situations such as free surface
avalanches, hopper flows, and boundary driven flows
\cite{midi2003,pouliquen2002}.  These flows are often associated with
large velocity gradients localized to a small region of space, with
little shear elsewhere.  Boundary driven shear flows have been
extensively studied in Couette geometries in both two and three
dimensions
\cite{veje1999,howell1997,howell1999,utter2004,mueth2000,mueth2003,losert2000,losert2001,khosropour1997}.
In these experiments, the shear band is localized near the inner
cylinder wall, regardless of which cylinder is rotating, because the
stress is larger at the inner cylinder as a result of the curvature of
the system \cite{losert2001}.

Shear flows have been studied in other geometries such as the modified
Couette cell of Van Hecke {\em et al.}
\cite{fenistein2003,fenistein2004}, yet curvature effects continue to
effect the nature of the shear band \cite{unger2004}.  In annular
shear cells, shear is induced between parallel plates on the top and
bottom of the packing, but the presence of gravity and stationary side
walls breaks the symmetry between the surfaces, and causes a shear
band to form at the top surface \cite{savage1984,tsai2003,tsai2004}.
Simulations of planar shear in the presence of a gravitational field
have seen similar behavior, with the bottom portion of the pack at
rest and a flowing region near the top surface
\cite{thompson1991,zhang1992}.

A theoretically simpler system consists of shear induced between two
parallel plates.  Any resulting spatial inhomogeneities must arise
from self-organization rather than from the asymmetries of the
driving.  Simulations of such shear have found both fluid-like and
solid-like behavior \cite{aharonov2002,babic1990}, and in
Lennard-Jones liquids are found to depend strongly on the boundary
conditions \cite{thompson1990}.  Many theories of granular shear
assume homogeneous driving, yet nearly all experimental studies have
been unable to achieve this ideal.  An experimental system capable of
large scale shear flows in a truly parallel plate geometry has been
lacking.

In this paper, I describe a novel parallel plate shear cell.  The
system is designed to contain the granular pack within a rectangular
well.  Shear is applied by controlling the motion of the front and
back walls.  The side walls are deformable in order that they do not
disturb the motion of the grains.  With this parallel plate shear
cell, I performed experiments on the initiation and evolution of shear
bands as a function of the shearing history of the granular pack.  I
present results which demonstrate a dependence of the shear profile on
the amount and manner in which the pack has been sheared in the past.
For amorphous packs, which have not been previously sheared, the
initial velocity profile is very wide with large variations from run
to run.  As the system is sheared further, the velocity profile
evolves to a sharply sheared form.  Upon reversal of the direction of
shear, there is a sudden change in the velocity profile, attributed to
the breaking and reforming of the contact-force network in agreement
with recent work in two-dimensional Couette studies by Utter and
Behringer \cite{utter2004} and in three-dimensional Couette flow
experiments by Losert and Kwon \cite{losert2001}.

I also explored the effects of cyclic shear, finding a dramatic
interplay between the spatial localization of shear flows and the
structural organization of the grains.  After only a few cycles of
back and forth shearing of a packing of monodisperse glass spheres, a
layered structure begins to form within the granular pack.  The
velocity profile across these layers shows jumps in the location of
the largest velocity gradient.  Periods of large slip between layers
are interspersed with periods of very little slip, during which the
shear is localized at one of the boundaries of the layered region.

\section{Background}

Several experiments on granular shear flow have been performed in a
Couette geometry.  Typically, when measuring the time-averaged
azimuthal velocity in steady-state flows, one finds that the velocity
decays very rapidly with radial distance from the inner cylinder
\cite{veje1999,bocquet2002,howell1997,howell1999,utter2004,mueth2000,mueth2003,losert2000,losert2001,khosropour1997}.
Such velocities are often measured by applying particle tracking
techniques to the motion of the individual grains at the boundaries of
three-dimensional packs or in the bulk of two-dimensional samples.
The velocity profile at the top surface of three-dimensional Couette
systems was found to decay with distance from the inner cylinder with
a form that was between exponential and Gaussian in shape
\cite{bocquet2002,losert2000}.  Bulk measurements in three dimensions
were made by Mueth {\em et al.}  using MRI techniques to track the
average mass flow \cite{mueth2000}.  They found little difference
between flows within the pack at different heights, and the flows at
the bottom surface as visualized through high-speed video.  For
samples of non-spherical, rough, or polydisperse grains, the azimuthal
velocity was found to decay as a Gaussian, $v(r) = v_0
\mathrm{exp}(-c({r \over d} - {r_0 \over d})^2$, with bead diameter,
$d$, and the center, $r_0$, located near the inner cylinder wall.
Packings of smooth monodisperse spheres formed layers, in which case
the profile was dominated by slip between these layers leading to a
nearly exponential profile, $v(r) = v_0 \mathrm{exp}({-br \over d})$.
In each case, the steady-state velocity profile, when normalized to
the applied shear rate, ($v_\mathrm{normalized}(r) = {v(r) \over
v(r=0)}$, where $v(r=0)$ is the speed of the inner cylinder) was
independent of the magnitude of the shear rate for slow flows.

At higher shear rates, Losert and Kwon found a roughly exponential
profile with a width that increased weakly with shear rate
\cite{losert2001}.  In two-dimensional experiments, the global packing
fraction is fixed, and is shown to have a large effect on the shear
flows \cite{veje1999}.  Above a critical packing fraction, the velocity
profile is again found to be exponential, and approximately
independent of the shear rate.   It should be noted, however, that
molecular dynamics simulations of boundary-driven planar shear have
found a critical velocity for the development of nonlinear velocity
profiles based on dilation effects \cite{xu2004}.

Fenistein {\em et al.} modified the Couette geometry to have a split
bottom surface \cite{fenistein2003,fenistein2004}.  In this way a
shear band is created in the bulk of the pack, away from the side
walls of the container.  As the depth of the pack is increased, the
width of the shear band grows and the position shifts toward the
center of the system.  The profile resembles a smooth transition
between two regions of nearly constant velocity.  An error function is
found to describe this profile for all conditions, $v(r) = {1 \over 2}
+ {1 \over 2} \mathrm{erf}({r-r_0 \over W})$.

Many theoretical models have been proposed to describe these flows.
Hydrodynamic models have been introduced with a density dependent
viscosity to account for dilation and localization of shear
\cite{losert2000,lubensky2003,santos2003}.  Debregeas and Josserand
have constructed a self-similar model based upon the intermittent
motion of various sized clusters of particles \cite{debregeas2000},
and \AA str\"om {\em et al.} have considered local rotating bearings
inside shear bands \cite{astrom2000}.

In a two dimensional Couette cell it is possible to measure the flow
patterns and the contact forces between particles simultaneously.
Howell {\em et al.} observed inhomogeneous force networks and their
fluctuations during shear using photoelastic disks
\cite{howell1997,howell1999}.  Utter and Behringer then used this
system to explore transients in the force network \cite{utter2004}.
As the system is sheared, the force network aligns at $45$ degrees to
resist the shear.  When the system is stopped these forces relax
somewhat, but upon restarting the shear in the same direction, there
is an immediate return to steady-state behavior.  However, when the
system is stopped and restarted in the opposite direction, the
velocity profile is initially much wider.  This is associated with
breaking of the previously aligned force network and reforming it in a
perpendicular orientation to resist the new direction of shear.
Similar transients in the velocity profile upon reversing the
direction of shear were seen previously in the three-dimensional
Couette studies of Losert and Kwon \cite{losert2001}.

In addition to these short-time transients, long-time changes in
behavior have been seen in granular flows.  Tsai and Gollub have shown
that boundary induced shear of a packing of monodisperse spheres can
lead to spatial ordering of the particles in an annular shear cell (in
which a vertical shear is applied between the floor and ceiling)
\cite{tsai2003,tsai2004}.  When shearing continuously under identical
conditions, some experimental runs found a crystallization transition,
at which point the sample became highly ordered.  Other runs led to a
compacted but still disordered final state.  By applying cyclic shear
to the packing before beginning their continuously shearing run, they
were able to select for the final crystallized state.  It is still
unclear how cyclic shear affects the ordering and whether it is more
efficient at creating order than unidirectional shear.  In other
geometries, cyclic shear has been shown to compact and crystallize
granular packings rapidly \cite{pouliquen2003}.

Shear has been seen to induce order in a wide variety of systems.
Shear induced alignment has been seen in liquid crystals
\cite{golan2001} and thin liquid films \cite{drummond2002}.  Thompson
{\em et al.} have seen shear-induced ordering in fluid film
simulations and an associated transition from the normal Newtonian
response \cite{thompson1995}.  In experiments on colloidal
suspensions, Cohen {\em et al.}  have shown that shear, and the
geometry in which it is applied, has a large impact on the spatial
ordering of particles \cite{cohen2004}.  Simulations by Stevens and
Robbins have shown that shear can effect the solid-liquid phase
boundaries of particles with screened Coulomb interactions, showing
instances of shear-induced melting and shear-induced ordering
\cite{stevens1993}.

Paulin {\em et al.} and Haw {\em et al.} have seen spatial ordering in
dense colloidal suspensions in response to both unidirectional and
cyclic shear \cite{paulin1997,haw1998_1}.  They found that the
particles order into hexagonal-close-packed layers.  Under some
instances a random stacking of these layers was produced with the
closest-packed direction oriented along the direction of flow.  Under
shear these layers were able to slide past each other.  Other
instances resulted in close-packed layers which were stacked in an
ABCABC... pattern of a three-dimensional face-centered-cubic crystal.
In these cases the closest-packed direction is oriented along the
direction of the velocity gradient (perpendicular to the direction of
flow), in which case only a small amount of shear could be sustained
within the crystal before disrupting the microstructure.

\section{Experimental Methods}

\begin{figure}[t] 
\begin{center}
\includegraphics[width=8.5cm]{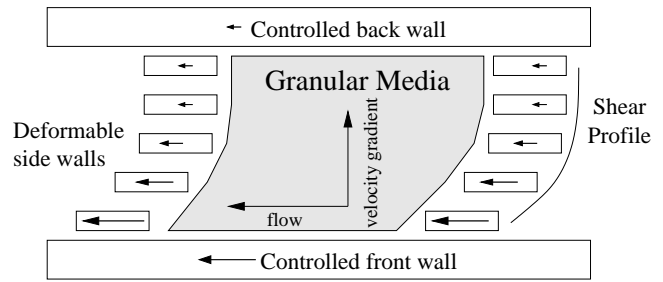}
\end{center}
\caption[Top view diagram of the parallel plate shear cell]{Top view
diagram of the parallel plate shear cell.  The front and back walls
induce shear across the granular pack.  The side walls deform in
response to the motion of the grains and allow for the recording of
the velocity profile of the granular material.}
\label{top_view}
\end{figure}

It is possible to contain granular materials within a
rectangular-shaped box and apply shear by moving the front and back
walls. Unfortunately, under any significant amount of shear, the
presence of the side walls disturbs the motion of the grains and
affects the shear profile.  A better situation is found when the side
walls are made to be deformable.  In this way the front and back walls
apply the shear to the granular pack, while the side walls contain the
pack, but do not overly disturb its motion.  Deformable side walls
also provide a unique opportunity to measure the motions of the grains
easily.  Recording the motions of the side walls provides an average
of the motions of the grains at that distance from the front and back
walls.  Figure \ref{top_view} shows a schematic of the motion of the
deformable side walls in response to a shear imposed by the front and
back walls.

\begin{figure}[t] 
\begin{center}
\includegraphics[width=7.5cm]{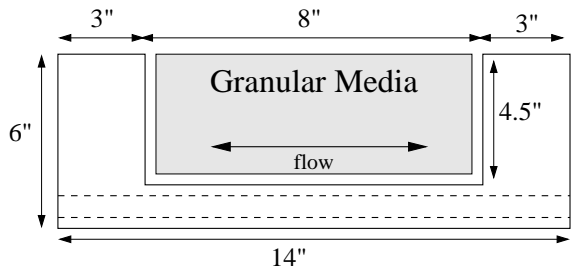}
\end{center}
\caption[Side view diagram of one slat]{Side view diagram of one slat.
Many such slats are stacked together to create a well in which to
contain the granular material.  The dashed lines represent two
v-grooves machined into one side of each steel slat to hold two rows
of ${1 \over 8}$ inch ball bearings which separate neighboring slats.
Each slat is free to slide along its length in response to the flow of
the grains.  The velocity gradient of the granular pack is
perpendicular to the page.}
\label{side_view}
\end{figure}

\begin{figure}[t] 
\begin{center}
\includegraphics[width=8.5cm]{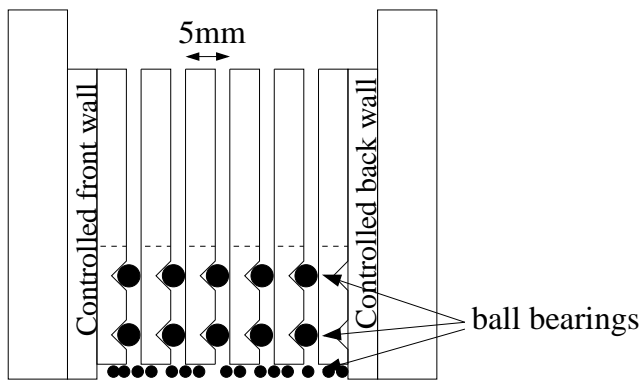}
\end{center}
\caption[End view diagram of the parallel plate shear cell]{End View
diagram of the parallel plate shear cell. Neighboring slats are
separated by two rows of ${1 \over 8}$ inch ball bearings.  Smaller
ball bearings separate all slats from the bottom floor.  The front and
back walls are moved into and out of the page in order to apply the
shear.  All other slats are independently movable, and respond to the
motion of the grains by moving in the direction perpendicular to the
page.}
\label{end_view}
\end{figure}

The cell is composed of many independently movable slats.  As the
grains move, these slats move past one another, adjusting the shape of
the confining well.  Figure \ref{side_view} shows one such slat (made
of ${1 \over 8}$ inch thick polished steel), with the granular pack
residing in the well.  One side of each slat contains two v-grooves,
$0.075$ inches deep, running parallel to the bottom surface.  Ball
bearings placed in these grooves separate neighboring slats by
approximately $2$mm, giving a repeat spacing between slats of $5$mm,
as shown in figure \ref{end_view}. Additional $1.5$mm ball bearings
reside underneath all of the slats, separating the system from the
floor of the setup.  This system of ball bearings provides low
friction between neighboring slats and between each slat and the
floor.  Each slat is able to respond to the motion of the grains,
without being overly influenced by the motions of the other slats.

Flat front and back walls are used to apply shear to the granular
pack. A single dilute layer of $5$mm glass beads is epoxied to the
central region of each of the front and back walls, in order to
provide a rough surface from which to induce shear.    The well is
$114$ mm deep and $203$ mm long. When using $5$mm beads this
means the granular pack is approximately $23$ bead diameters deep and
$41$ bead diameters long.  Multiple slats are stacked together to
control the width of the well.  The data presented here was taken with
$24$ slats, with a repeat spacing similar to a grain diameter,
resulting in a pack that is approximately $24$ bead diameters wide.

In order to test the reliability of the parallel plate shear cell,
multiple control experiments were conducted.  When the system was
filled with granular material and both front and back walls were
driven in the same direction at the same speed, the entire granular
pack and all intermediate slats moved together at this same speed.
The friction with the bottom floor was not sufficiently large to
impede the motion of the individual slats.

Without any granular material residing in the well, each individual
slat was pushed by hand while the other slats were watched for
motion.  Neighboring slats were not dragged along with the motion of
any given slat.  This confirms that the friction between slats is
small, and that each slat is independently movable and capable of
responding to the motions of the granular pack.  This test is repeated
before and after each run.  Data was taken with different sized
particles, at different absolute speeds and with different speed
ratios between the front and back walls, with either the front or back
wall being faster.  The system was started at various positions and
the slats were started in various configurations with no qualitative
difference in results.

The front and back walls can be driven at varying speeds (from
approximately $0.5$ mm/s to $3$ mm/s) in either direction.  If the
walls are driven in opposite directions, there will necessarily be a
region of the granular pack at rest with respect to the bottom floor.
In practice, the static friction of the slats with the floor tends to
inhibit the motion of the central slats, resulting in regions of large
shear at both walls, and little shear in the center of the pack.  For
the data presented here, both walls were driven in the same direction
so that the entire system moved at least as fast as the slower wall.
In this way no slat is at rest with respect to the floor.

Initially the $5$mm glass beads are poured into the well created by
the slats and mixed by hand to disrupt any ordering.  A single DC
motor drives both walls through ACME lead screws.  The system is
videotaped with a digital video camera.  Individual video frames
are analyzed with IDL \cite{idl} to locate the positions of the
leading edges of each slat.  These positions are tracked as a function
of time and analyzed to give the distance that each slat has
moved and it's velocity as a function of time.

\section{Results}

\begin{figure}[t]
\begin{center}
\includegraphics[width=8.5cm]{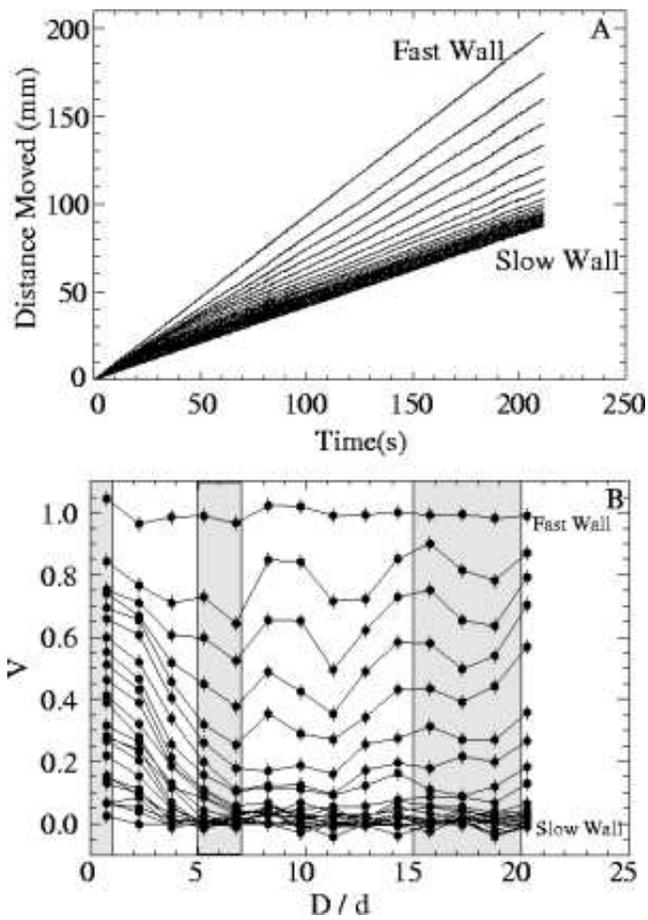}
\end{center}
\caption[Distance moved versus time and scaled velocity, $V$, versus
distance sheared, $D$.]{a) Distance that each slat has moved as a
function of time for one experimental run.  The top line represents
the fast moving wall, and the bottom line represents the slow moving
wall.  b) Scaled velocity, $V$, versus distance sheared, $D$, in units
of a bead diameter, $d$.  Each data point was obtained from a linear
fit to the positions of each slat as a function of time over a window
of approximately $14$s, or $1.5$ bead diameters of shear.  Shaded
regions show time windows which are used to compute velocity profiles
in figures \ref{initial_profiles} and \ref{profiles_mid_and_late}.}
\label{time_and_vel_traces}
\end{figure}

Beginning with an amorphous pack of monodisperse spheres, I track the
positions of each of the slats, and thus the velocity profile of the
granular pack.  Figure \ref{time_and_vel_traces}a shows the distance
that each slat has moved as a function of time for one experimental
run.  The steepest line represents the fast wall, and the shallowest
line the slow wall.  Intermediate slats move at speeds between that of
the slow wall and the fast wall.

Applying a linear fit to each slat position versus time over some time
window allows us to measure the velocities of each slat.  A small time
window allows good time resolution when looking for variations in the
velocity, but also results in a large uncertainty in the fitted
velocity.  Throughout this analysis, the size of the time window is
adjusted as appropriate to each situation.  In order to compare flow
profiles for different experimental conditions, each slat velocity is
rescaled to make the slow wall have an average velocity of $0$ and the
fast wall have an average velocity of $1$.  For each slat, the scaled
velocity $V$, as a function of time is given by

\begin{equation}
V(t) = \frac {v(t) - v_\mathrm{slow}}{v_\mathrm{fast} -
v_\mathrm{slow}}
\label{scaled_velocity}
\end{equation}

where $v_{\mathrm{slow}}$ and $v_{\mathrm{fast}}$ are the average
velocities of the slow and fast walls respectively.  Time is scaled by
the relative velocity of the front and back walls, so as to represent
the amount of relative displacement of the fast wall with respect to
the slow wall.  This distance, $D$, is referred to as the distance
sheared.

\begin{equation}
D = t (v_{\mathrm{fast}} - v_{\mathrm{slow}})
\end{equation}

\begin{figure}[t] 
\begin{center}
\includegraphics[width=8.5cm]{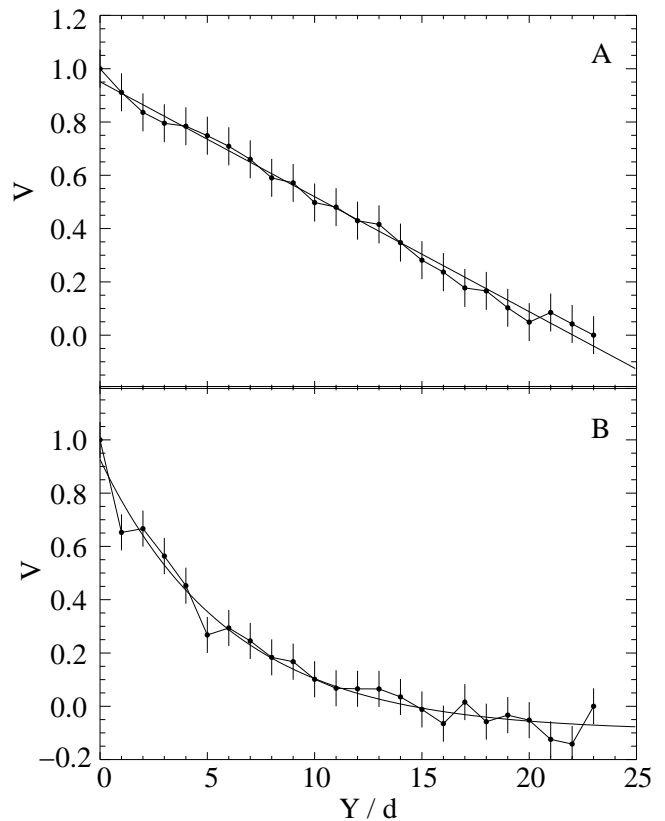}
\end{center}
\caption[Initial velocity profiles]{Two extreme examples of the range
of possible velocity profiles when shear is first initiated on an
amorphous granular pack.  Scaled velocity, $V$, is plotted versus
distance from the fast wall, $Y$, in units of a bead diameter, $d$.
In some cases the profile is linear across the entire width of the
setup, as shown in part a.  In other instances the profile is sheared
more sharply near the faster wall as shown in part b.  The solid line
shows an exponential fit resulting in  $1.0 \mathrm{exp}(\frac
{-Y}{6.1d}) - 0.09$.   Velocities were taken from linear fits to the
positions of each slat as a function of time over the first $1.0$ bead
diameters of shear, marked as the first shaded region in figure
\ref{time_and_vel_traces}b.}
\label{initial_profiles}
\end{figure}

Figure \ref{time_and_vel_traces}b shows the scaled velocity, $V$, of
each slat versus distance sheared, $D$.  The absolute velocity is not
found to have a qualitative effect on the flow profile.  When shear is
first initiated across an amorphous packing of glass beads, the
velocity profile varies significantly from run to run.  The first
column of points in figure \ref{time_and_vel_traces}b shows a rather
broad profile, which is a typical response.  However, This initial
velocity profile has been found to range from a linear profile across
the entire width of the setup to an exponential profile with a width
as small as $6$ bead diameters.  Figure \ref{initial_profiles} shows
two extreme examples of the scaled velocity, $V$, versus distance from
the faster wall, $Y$.  When the system shows a nonlinear profile, then
there always exists more shear near the faster moving wall (regardless
of whether this is the front wall or the back wall).

\begin{figure}[t]
\begin{center}
\includegraphics[width=8.5cm]{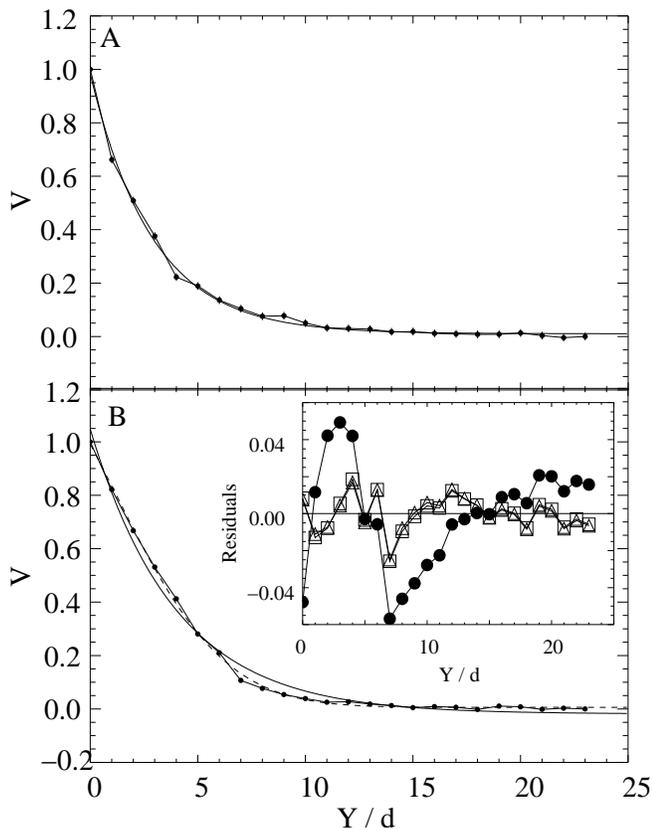}
\end{center}
\caption[Scaled velocity profiles after varying amounts of shear]{a)
Scaled velocity, $V$, versus distance from the fast wall, $Y$, after
$5$ to $7$ bead diameters of shear (see second shaded region of figure
\ref{time_and_vel_traces}b).  The solid line is an exponential fit,
resulting in $0.97 \mathrm{exp}(\frac {-Y} {2.9d}) + 0.01$.  b) Scaled
velocity, $V$, versus distance from the fast wall, $Y$, after $15$ to
$20$ bead diameters of shear (third shaded region in figure
\ref{time_and_vel_traces}b). The solid line is an exponential fit
($1.1 \mathrm{exp}({-Y \over 4.0d}) - 0.02$), which does not fit the
data well.  The dashed line is the result of a Gaussian fit ($1.3
\mathrm{exp}(({Y+3.9d \over 7.1d})^2) + 0.006$), which fits
significantly better.  An error function fit resulting in $-0.86
\mathrm{erf}({Y - 0.81d \over 6.0d}) + 0.86$ is indistinguishable from
the Gaussian at this scale.  The inset shows the residuals to the
exponential fit (filled circles), Gaussian fit (open squares), and
error function fit (open triangles).}
\label{profiles_mid_and_late}
\end{figure}

Despite the initial variation in the velocity profile, the profile
evolves to a sharper decay at the faster moving wall, as the system is
sheared.  Figure \ref{time_and_vel_traces}b shows that the velocities
of all but the slats closest to the fast wall quickly decay to nearly
as slow as the slow wall.  Figure \ref{profiles_mid_and_late}a shows
the velocity profile of this run fitted over the time from $5$ to $7$
bead diameters of shear (the second shaded region in figure
\ref{time_and_vel_traces}b).  The profile is sharply sheared near the
fast wall, exhibiting an exponential profile with a width of
approximately $3d$.  This trend is qualitatively reproduced for all
experimental runs, although the amount of shear required for this
evolution can vary from approximately $3$ to approximately $15$ bead
diameters of shear across the system.

As the system continues to shear, the profile does not get any sharper
than that shown in figure \ref{profiles_mid_and_late}a.  However, the
velocity profile does begin to lose its exponential character.  Figure
\ref{profiles_mid_and_late}b shows the velocity profile over the time
from $15$ to $20$ bead diameters of shear (see the third shaded region
of figure \ref{time_and_vel_traces}b).  The solid line shows a fit to
an exponential which systematically deviates from the data.  The inset
of figure \ref{time_and_vel_traces}b shows the residuals of the
exponential fit.  A better fit is obtained from a Gaussian function or
an error-function which is indistinguishable from the tail of a
Gaussian at this scale.  It should be noted that earlier profiles,
which are well fit by an exponential form, can also be fit to the
extreme tails of Gaussian or error function forms.  The resulting fit
parameters put the centers of the Gaussian and error function profiles
very far behind the fast wall ($<-10d$) with very large widths, and
are thus considered unreasonable.  As the system continues to shear
further, the fitted center of the Gaussian and error-function forms
move closer to the faster wall.  After $15$ to $20$ bead diameters of
shear (the farthest unidirectional shear possible in this system), the
fitted center of the Gaussian and error function are at $4$ bead
diameters behind the fast wall and $0.8$ bead diameters in front of
the fast wall with widths of $7d$ and $6d$ respectively.  It is
unclear how far these parameters would evolve if the system was able
to continue to shear in one direction.

\begin{figure}[t] 
\begin{center}
\includegraphics[width=8.5cm]{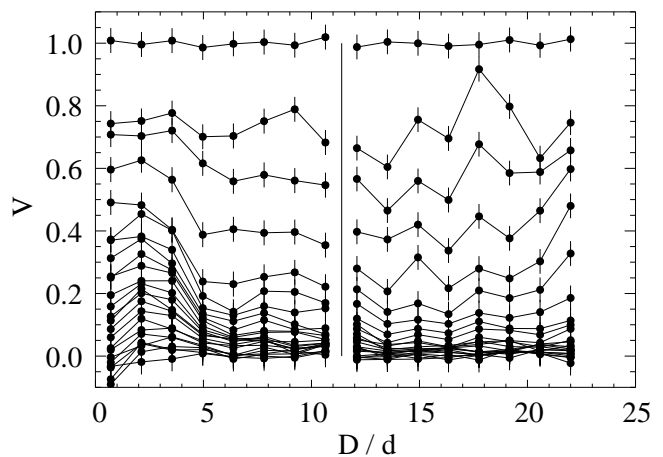}
\end{center}
\caption[Response to stopping and continuing in the same
direction]{Scaled velocity, $V$, versus distance sheared, $D$.  The
vertical line represents the time at which the system was stopped and
restarted in the same direction.  There is no significant change in
the velocities of the individual slats as a result of the stopping and
restarting.}
\label{vel_trace_break}
\end{figure}

\begin{figure}[t] 
\begin{center}
\includegraphics[width=8.5cm]{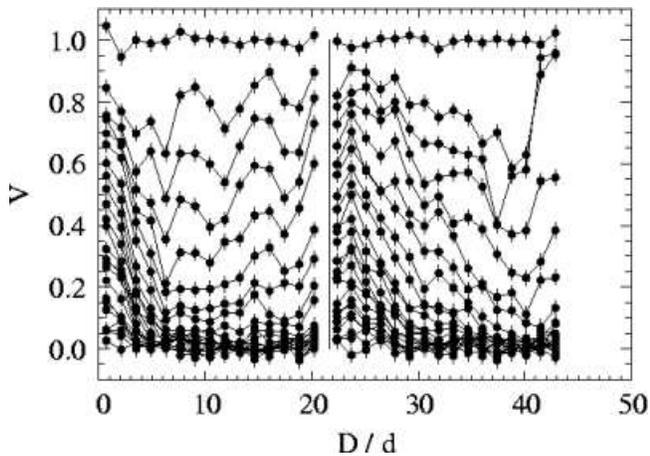}
\end{center}
\caption[Response to stopping and restarting in the reverse
direction]{Scaled velocity, $V$, versus distance sheared, $D$.  The
vertical line represents the time at which the system was stopped and
restarted in the opposite direction.  The magnitude of the speeds of
both the front and back walls remained the same, but both were now
moving in the reverse direction.  Immediately upon reversal of the
direction of shear, the velocity profile becomes much broader,
resembling the initial previously unsheared profile.  The system then
evolves again toward a more sharply sheared profile.}
\label{vel_trace_switch}
\end{figure}

If the system is stopped and restarted in the same direction, there is
little change to the velocity profile.  Figure \ref{vel_trace_break}
shows $V$ as a function of $D$ for a run which was stopped after
$11.5$ bead diameters of shear as denoted by the vertical line.  When
the system was restarted in the same direction the velocity of each
slat remained essentially unchanged.  By contrast, when the system is
stopped and restarted in the opposite direction, as shown in figure
\ref{vel_trace_switch}, there is a large and sudden change in the
velocity profile.  Immediately after reversal, the velocity profile
becomes much broader, resembling the initial profile of an unsheared
system.  The system again evolves toward a more sharply sheared
profile.

\begin{figure}[t] 
\begin{center}
\includegraphics[width=8.5cm]{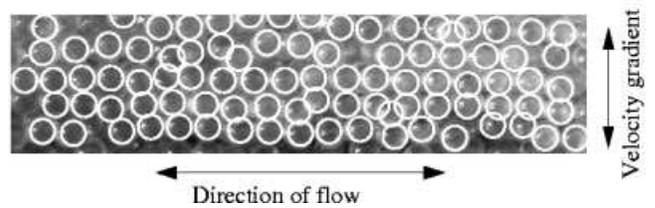}
\end{center}
\caption[Photograph of beginning of shear-induced layering]{A
photograph of the top surface of partial layering induced by cyclic
shear.  White circles have been added to highlight the
particles.}
\label{layers}
\end{figure}

After repeated cycling of the direction of shearing, an ordered
structure begins to emerge within the granular pack.  Layers form
parallel to the direction of flow, but there is little registration
between neighboring layers.  Figure \ref{layers} shows a picture of
the top surface of the packing.  By altering the amplitude of each
shear cycle, the region in which the layers form can be changed.
Typically they first appear at a distance from the faster wall where
there is approximately one bead diameter of shear per cycle across one
bead diameter of the packing (where the shear is approximately one).
Small amplitudes lead to the initial formation of layers near the fast
wall. Larger amplitudes form the layered structure farther from the
faster wall.

\begin{figure}[t] 
\begin{center}
\includegraphics[width=8.5cm]{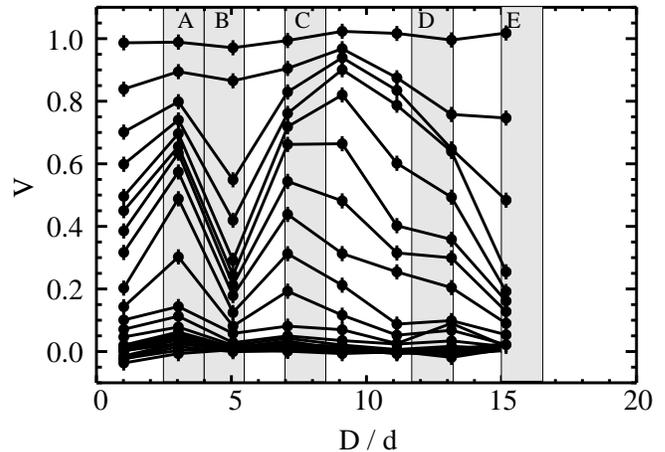}
\end{center}
\caption[Scaled velocity trace while layering]{Scaled velocity, $V$,
versus distance sheared, $D$, for a system which has previously
undergone $9$ cycles of forward and reverse shearing, each shearing
approximately $18$ bead diameters.  At this point the granular pack is
fairly well layered, but not completely crystallized in the region
from roughly $4$ bead diameters from the fast wall to roughly $8$ bead
diameters from the fast wall.  There exist periods of time in which
there is very little slip between layers in which case the shear is
localized at one of the boundaries of the layered region as well as
periods of time of extreme slipping between layers.  Figure
\ref{layered_profiles} shows velocity profiles for the periods of time
denoted by the five shaded regions.}
\label{vel_trace_layers}
\end{figure}

\begin{figure}[ht] 
\begin{center}
\includegraphics[width=8.5cm]{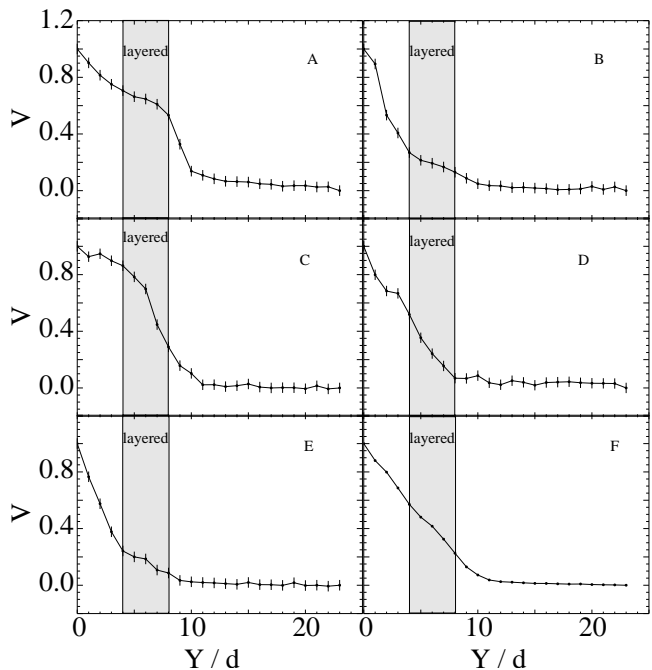}
\end{center}
\caption[Velocity profiles of partially layered system]{Scaled
velocity, $V$, as a function of distance from the faster wall, $Y$,
for a partially layered sample.  In the region from roughly $4$ bead
diameters from the fast wall to roughly $8$ bead diameters from the
fast wall (shaded) the granular pack is fairly well layered, but not
completely crystallized.  Parts a-e show the velocity profile when
shearing forward during the $10$th cycle from $2.5$ to $4$, $4$ to
$5.5$, $7$ to $8.5$, $11.7$ to $13.2$, and $15$ to $16.6$ bead
diameters respectively; these times are denoted by the five shaded
regions in figure \ref{vel_trace_layers}.  Part f shows the average
velocity profile across the entire $10$th cycle.}
\label{layered_profiles}
\end{figure}

As this layered structure is forming, the velocity profile undergoes
dramatic changes in terms of where the shear is localized.  Figure
\ref{vel_trace_layers} shows $V$ as a function of $D$ for a system
after it has already undergone $9$ complete forward and reverse cycles
of approximately $18$ bead diameters of shear in each direction.  At
different times the largest shear is located at different points
within or at the edges of the layered region.  After approximately $3$
bead diameters of shear (first shaded region), the shear is maximum
near the edge of the layered region farthest from the fast wall, which
is roughly $8$ bead diameters from the fast wall.  The velocity
profile at this point (shown in figure \ref{layered_profiles}a), shows
a small amount of shear between the fast wall and the layered region,
very little shear within the layered region, and a lot of shear at the
far edge of the layers.  As the system shears farther, the area of
maximum shear moves to the fast wall as shown in the profile in figure
\ref{layered_profiles}b.  In parts c and d, the shear is maximum
within the layered region, while at later times it moves back to the
fast wall (part e.)  Figure \ref{layered_profiles}f shows the average
velocity profile across the entire forward shearing of the $10$th
cycle and shows a fairly linear decay out to a distance of
approximately $11$ bead diameters from the fast wall and very little
shear beyond that.  Such a time-averaged profile, as has often been
done in other experiments for improved accuracy, is incapable of
capturing the true migration of the shear band.

\begin{figure}[ht] 
\begin{center}
\includegraphics[width=8.5cm]{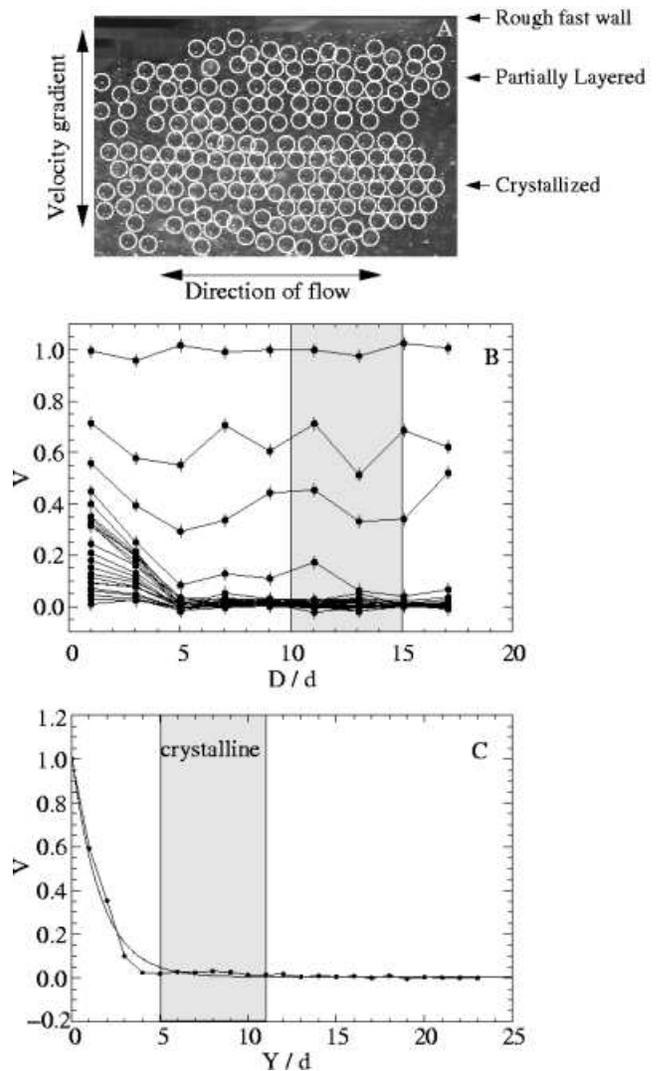}
\end{center}
\caption[Picture of crystallized region, velocity trace, and velocity
profile in a partially crystallized pack]{a) Shear-induced
crystallization.  A picture of the top surface shows a crystalline
region between approximately $5$ bead diameters and $12$ bead
diameters from the fast wall, with partial layering between this
crystalline region and the fast wall.  White circles have been
placed around particle images.  b) The scaled velocity, $V$, is
plotted versus distance sheared, $D$, showing an initial response to
changing direction and at later times showing a very sharp profile.
c) The scaled velocity, $V$, as a function of distance from the fast
wall $Y$, during the time from $10$ to $15$ bead diameters of shear,
as denoted by the shaded region in part b. There is virtually no shear
within the crystalline region.  The solid line is the result of an
exponential fit ($1.0 \mathrm{exp}({-Y \over 1.6d}) + 0.003$), which
is not sharp enough to capture the data.}
\label{crystal}
\end{figure}

Continuing to shear the system cyclically causes the layered region to
crystallize.  Figure \ref{crystal}a shows a sample which after $29$
cycles became very well ordered from approximately $5$ bead diameters
from the fast wall to approximately $11$ bead diameters from the fast
wall, consisting of horizontal hexagonal-close-packed layers with the
close-packed direction oriented parallel to the direction of flow.
The area between the fast wall and the crystal was somewhat layered,
but not completely crystalline, with a very sharp boundary marking the
edge of the crystal.  Deconstruction of the crystal, by vacuum removal
of individual grains, revealed that the structure continued downward
through the packing to the bottom surface.  Furthermore, it was found
that these close-packed layers were ordered in a ABCABC... pattern,
forming a three dimensional face-centered-cubic crystal.

Figure \ref{crystal}b shows $V$ as a function of $D$ for this
partially crystallized packing.  The initial profile upon starting
this $30$th shear cycle (after switching direction from the end of the
$29$th cycle) shows a broad profile, consistent with the same response
to switching direction that was seen in amorphous packings.  At later
times essentially all shear occurs within the partially layered region
between the fast wall and the crystalline area.  An exponential fit
gives a width of $1.6d$, but even this form is not sharp enough to
completely capture the data. Virtually no shear is found within the
crystal, resulting in a flat profile beyond $4$ bead diameters.

\section{Discussion}

The initial velocity profiles upon shearing an amorphous pack vary
significantly from packing to packing.  There is a tendency for the
system to shear more strongly at the faster moving wall, which might
be an indication of inertial effects, enhanced friction with the
bottom surface for the faster moving slats, or larger vibrations at
the faster moving wall.  This breaks the symmetry between the fast and
slow walls, and provides a preferential position for the formation of
a shear band.  However, the wide variation in initial profiles
suggests that this symmetry may be only weakly broken and that it
therefore requires a varying amount of time for the shear band to form
at the fast wall.  This implies that there is an initial period before
which, a shear band has not formed, and that this period can last
varying amounts of time depending upon the details of the packing.

The widely varying initial velocity profile evolves to a more sheared
profile (as shown in figure \ref{profiles_mid_and_late}a) with an
exponential width of approximately $3$ bead diameters.  After further
shearing, the profile consistently deviates from exponential.  This
profile is not affected by suddenly stopping and restarting the
shearing process.  As has been indicated by Utter and Behringer
\cite{utter2004} and by Losert and Kwon \cite{losert2001}, slow dense
granular flows are quasistatic.  All information regarding previous
shear history is stored in the inter-particle contacts, which are not
overly affected by stopping and restarting the flows
\cite{utter2004,losert2001}.  These contacts, however, are greatly
affected by reversing the direction of the flow.  Howell and Behringer
visualized the force network in two dimensional flows and found that
reversing the direction of flow causes the network to break and reform
in a perpendicular orientation \cite{utter2004}.  As the force network
is reforming, particles far from the shear band move significantly
more than in steady state, resulting in a much wider velocity profile.
I observe this effect in amorphous packings in agreement with Utter
and Behringer's two dimensional experiments and Losert and Kwon's
three-dimensional Couette experiments.  I also see a similarly wide
velocity profile when switching directions in layered and crystalline
packings.  This suggests that the formation of an aligned force
network is a function of the microscopic contacts between particles,
and that the time scale for such formations is not strongly affected
by macroscopic ordering of the particles.  Furthermore, the force
network can be broken and reformed without disrupting the large-scale
crystal structure.

Like Tsai and Gollub \cite{tsai2004}, I found that shear can induce
spatial ordering of granular particles.  The time scale for such
formation seems to be much faster under cyclic shear than under the
unidirectional shear of their experiments, requiring only a few tens
of cycles to create order over several particles.  However, it is not
clear that cyclic shear would be capable of crystallizing the entire
pack, as particles near the slow wall experience very little shear
within a given cycle.  The ordering is most pronounced at the position
in the packing where the shear per cycle is approximately one (where
there is approximately one bead diameter of shear across a width of
one bead diameter for each cycle).  By changing the amplitude of the
shear cycles I was able to form layered regions at different positions
within the packing.

Over the range of shear amplitudes from $1$ bead diameter to $20$ bead
diameters, all ordered regions formed layers and crystalline
structures oriented with the close-packed direction parallel to the
direction of flow.  I never observed the formation of order with the
close-packed direction parallel to the velocity gradient, as was seen
in the colloidal studies of Paulin {\em et al.} and Haw {\em et al.}
\cite{paulin1997,haw1998_1}.  I did, however, observe cases of both a
random stacking of planes in the layered packings and a vertical
ABCABC... pattern forming a face-centered-cubic (FCC) structure in the
crystalline packings.

As an ordered structure is beginning to form, the velocity profile
undergoes repeated changes between periods of time during which there
is large shear within the ordered region and periods of time where
there is virtually no shear within that region.  At later times, a
highly-ordered FCC structure emerges, in which there is no shear.  The
entire crystalline region moves as a solid block throughout the rest
of the experimental run.  This might be related to the unique way in
which FCC packings support stress.  Experiments on static granular
packings have shown that forces are supported along straight lines in
face-centered-cubic packings whereas in  other stackings of
hexagonally packed layers, such as in a hexagonal-close-packed
crystal, the forces branch or split between each close-packed plane
\cite{mueggenburg2002}.   Any out-of-plane force is supported by a
network which  branches at each horizontal layer of the packing that
deviates from a local FCC stacking order \cite{spannuth2004}.  These
branches can create forces which would cause a slipping of horizontal
planes at regions of the packing which deviate from the
ABCABC... stacking order.  In a face-centered-cubic crystal,
out-of-plane forces are supported by straight lines of force across
the entire packing without branching, and without in-plane forces to
encourage slipping.  Thus a connection can be made between the force
network within a packing and the response of that packing to shear.
As a shear-induced ordered structure is formed, local regions of
non-FCC stacking order experience slip between the horizontal planes,
and reorder, while regions of FCC stackings, are more stable to
rearrangments.  Over long times, the ordered region develops a large
scale three-dimensional face-centered-cubic structure which moves
together as a solid block.  This may be a mechanism whereby
long-range three-dimensional order in a granular pack can be created
by shear.  Further  tests need to be preformed to confirm that FCC
packings are more stable to shear than other stacking orders, and that
this stability can be related to the force networks within the
packings.

It should be noted that my results show velocity profiles similar in
character to those found by Aharonov and Sparks in two-dimensional
simulations \cite{aharonov2002}.  However, Aharonov and Sparks found
switching between randomly positioned internal shear bands and diffuse
deformations in the regime of large applied pressures, and found the
more standard narrow shear band at a boundary under smaller confining
pressures.  I find these two different regimes of shear flow based
upon the structural order of the packing.  It is unclear if a
connection could be made comparing the microstructural reorganization
from applied pressures and from shear-induced ordering.

\section{Conclusion}

In this paper I have presented the design and implementation of a
novel parallel plate shear cell for the purpose of studying cyclic
granular shear.  The parallel plate geometry avoids the curvature
effects found in the more common cylindrical Couette experiments.  The
system of adjustable side walls allows for reasonable distances of
shear in one direction, and is especially well suited for cyclic shear
studies.

Experiments in this setup were done on the shear flows of packings of
monodisperse glass spheres, studying transient behaviors, and
shear-induced crystallization.  There is a complex interaction between
the spatial structure and the shear flow properties of a granular
pack.  The initial incomplete formation of layers can allow for more
shear than in an amorphous pack through slipping between layers.  Yet
the layers can also lock together, especially as they become more well
ordered, preventing slip and resulting in less shear in that region.
This nontrivial dependence of shear flows on the amount and type of
structure within the granular pack signifies the complex dynamics
inherent in such non-equilibrium systems.

\begin{acknowledgments}
My utmost gratitude is given to Heinrich Jaeger and Sidney Nagel for
their invaluable guidance and support.  I also thank B. Chakraborty,
X. Cheng, E. Corwin, M. M\"obius, and T. Witten for fruitful
discussions.  This work was supported by the NFS under CTS-0090490 and
MRSEC DMR-0213745, and by DOE under W-7405-ENG-82.
\end{acknowledgments}

\end{document}